\documentclass[conference,twoside]{csce}

\usepackage[hmargin=.75in,vmargin=1in]{geometry}
\usepackage[american]{babel}
\usepackage[T1]{fontenc}
\usepackage{times}
\usepackage{caption}
\usepackage{balance}
\usepackage{fancyhdr}
\usepackage{textcomp}
\usepackage{epsfig,graphicx}
\usepackage{xcolor}
\usepackage{amsfonts,amsmath,amssymb}
\usepackage{fixltx2e} 
\usepackage{booktabs}

\renewcommand{\thispagestyle}[2]{} 
\title{\bf Shed More Light on Bloom Filter's Variants}           

\author{
{\bfseries Ripon Patgiri, Sabuzima Nayak, and Samir Kumar Borgohain}\\
Department Of Computer Science \& Engineering, National Institute of Technology Silchar, Assam, India
}

\begin{document}

\maketitle                        

\begin{abstract}
Bloom Filter is a probabilistic membership data structure and it is excessively used data structure for membership query. Bloom Filter becomes the  predominant data structure in approximate membership filtering. Bloom Filter extremely enhances the query response time, and the response time is very fast. Bloom filter (BF) is used to detect whether an element belongs to a given set or not. The Bloom Filter returns \textbf{True Positive (TP)}, \textbf{False Positive (FP)}, or \textbf{True Negative (TN)}. The Bloom Filter is widely adapted in numerous areas to enhance the performance of a system. In this paper, we present a) in-depth insight on the Bloom Filter,and  b) the prominent variants of the Bloom Filters.
\end{abstract}

\vspace{1em}
\noindent\textbf{Keywords:}
 {\small  Bloom Filter, Scalable Bloom Filter, Variants of Bloom Filter, Membership filter, Data Structure, Algorithm} 


\section{Introduction}
The Bloom Filter \cite{Bloom} is the extensively used probabilistic data structure for membership filtering. The query response of Bloom Filter is unbelievably fast, and it is in $O(1)$ time complexity using a small space overhead. The Bloom Filter is used to boost up query response time, and it avoids some unnecessary searching. The Bloom Filter is a small sized data structure. The query is performed on the Bloom Filter before querying to the large database. The Bloom Filter saves immense query response time cumulatively. However, there is also a false positive which is known as overhead of the Bloom Filter. Nevertheless, the probability of the false positive is very low. Thus, the overhead is also low. Moreover, a careful implementation of Bloom Filter is required to reduce the probability of false positive. 

There are various kind of Bloom Filters available, namely,  Blocked Bloom Filter \cite{Putze}, Cuckoo Bloom Filter \cite{Cuckoo}, d-Left CBF (dlCBF) \cite{d-left}, Quotient Filter (QF) \cite{Bender}, Scalable Bloom Filter (SBF) \cite{Almeida}, Sliding Bloom Filter \cite{Naor}, TinySet \cite{TinySet}, Ternary Bloom Filter (TBF) \cite{Lim}, Bloofi \cite{Adina}, OpenFlow \cite{openflow}, BloomFlow \cite{Craig}, Difference Bloom Filter (DBF) \cite{Yang}, and Dynamic Reordering Bloom Filter \cite{DChang}. The variants of Bloom Filters are designed based on the requirements of the applications. The Bloom Filter data structure is highly adaptable. Therefore, the Bloom Filter has met a enormous applications. A careful adaptation of the Bloom Filter ameliorates the system. However, it depends on the requirements of the applications. Bloom Filter's improvement potentiality makes the vast applicability of the probabilistic data structure. 


The fast query response using Bloom Filter attracts all the researchers, developers, and practitioners. There are tremendous applications of Bloom Filter. For instance, the BigTable uses Bloom Filter to improve disk access time significantly \cite{BigTable}. Moreover, the Metadata Server is drastically enhanced by Bloom Filter \cite{Anitha15,Zhu04,Zhu08,Hua11}. The Network Security is also boosted up using Bloom Filter \cite{zhu,mesh}. In addition, the duplicate packet filter is a very time consuming process. The duplicate packets are filtered in O(1) time complexity using Bloom Filter \cite{Fernandez}. Besides, there are diverse applications of Bloom Filter which improve significantly the performance of a system. The Bloom Filter predominant the filtering system, and thus poses some research questions (RQ) which are listed below-

\begin{description}
\item[RQ1:] Where should not Bloom Filter be used?
\item[RQ2:] What is the barrier of Bloom Filter?
\item[RQ3:] What are the various kinds of Bloom Filters available?
\end{description}

The research question (RQ) leads the article to draw a suitable conclusion. The RQ1 exposes the  reason for using Bloom Filter. The RQ2 exploits the False Positive of Bloom Filter. And finally, the RQ3 exposes the state-of-the-art development of Bloom Filter.


\section{Bloom Filter}
\label{bloomfilter}
\begin{figure}[ht]
\centering
\includegraphics[width=0.4\textwidth]{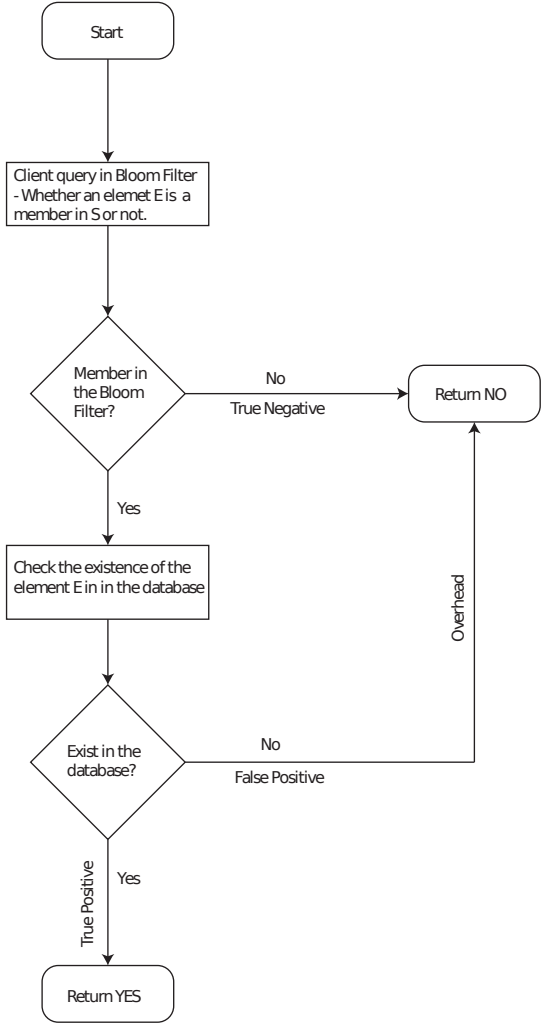}
\caption{Flowchart of Bloom Filter. Figure demonstrates the overhead of Bloom Filter}
\label{BF}
\end{figure}

The Bloom Filter \cite{Bloom} is a probabilistic data structure to test an element membership in a set \cite{Grandi}. The Bloom Filter uses a small space overhead to store the information of the element set. The True Positive and True Negative enhance the performance of filter mechanism. However, there false positive overhead in the Bloom Filter variants. However, the probability of false positive is negligible. But, some system cannot tolerate False Positive, because the false positive introduces error to the system. For example, duplicate key filtering system. Moreover, it also guarantees that there is no False Negative (FN) except counting variants of Bloom Filter. There are many systems where most of the queries are TN. Let $K=k_1,k_2,k_3,\ldots,k_n$ be elements present in the set $S$. Let $k_i$ be the random element where $1\leq i$. The approximate membership query is whether $k_i\in S$ or not.  The Bloom Filter returns either positive or negative. The positive is classified into False Positive (FP) and True Positive (TP). The FP of Bloom Filter returns existence of an element in a set, but $k_i\not\in S$. However, the TP correctly identifies the element, and it is in the set. Similarly, the negative is also classified into TN and FN. The TN boosts up the performance of a system and FP degrades the performance of a system. Therefore, the key challenge of Bloom Filter design is to reduce the probability of FP. The Figure \ref{BF} depicts the flowchart of Bloom Filter. The Figure \ref{BF} clearly exposes the overhead of Bloom Filter in case of FP.

\begin{figure}[ht]
\centering
\includegraphics[width=0.4\textwidth]{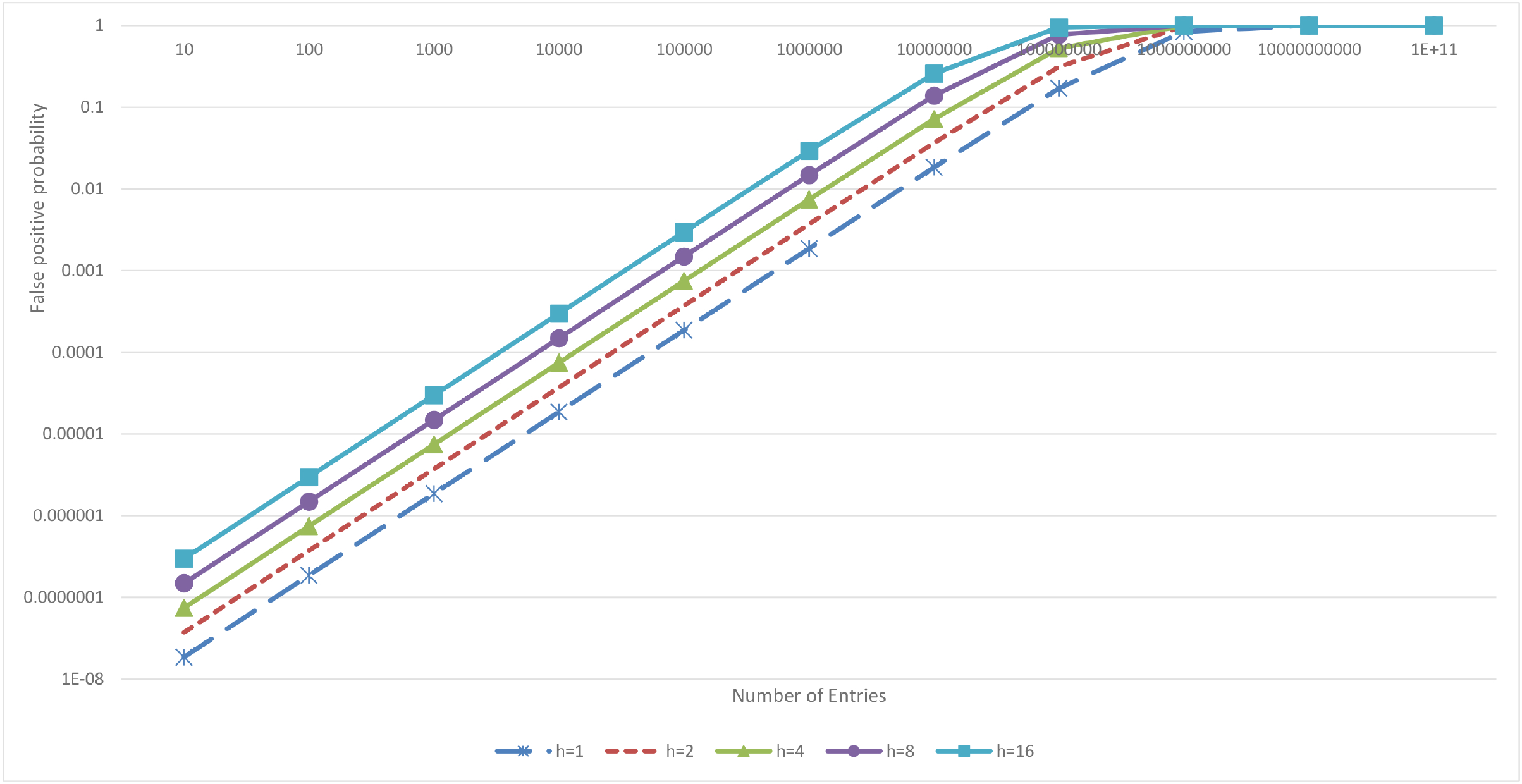}
\caption{The false positive probability of Bloom Filter with $m=64MB$, and $h=1,~h=2,~h=4,~h=~8,~and~h=16$. The X-axis represents the number of entries $n$.}
\end{figure}

\subsection{Analysis}
The FP affects on performance of Bloom Filter and this is an overhead of a system as shown in Figure~\ref{BF}. However, almost all variants of the Bloom Filter reduce the FP probability. Let us assume, $m$ is the number of bits available in the array. The probability of particular bit to be $1$ is $\frac{1}{m}$. The probability of particular bit to be $0$ is \[\left(1-\frac{1}{m}\right)\] Let $h$ be the number of hash functions and the probability of that bit remain 0 is \cite{Tarkoma,Grandi} \[\left(1-\frac{1}{m}\right)^h\] There are total $n$ element insertion into the array, therefore, the probability of that bit still $0$ is \[\left(1-\frac{1}{m}\right)^{nh}\] Now, the probability of that particular bit to be $1$ is \[\left(1-\left(1-\frac{1}{m}\right)^{nh}\right)\] What is the optimal value of hashing $h$? The probability of all bits 1 is \[\left(1-\left(1-\frac{1}{m}\right)^{nh}\right)^h\approx\left(1-e^{-hn/m}\right)^h\] The probability of false positive increases with the large size of entries $n$. However, it is reduced by increasing the value of $m$. Therefore, minimizing the false positive probability is \[h=\frac{m}{n}ln 2\] Let us $p$ be the desired false positive, and hence, \[p=\left(1-e^{-(\frac{m}{n}ln 2n)/m}\right)^{(\frac{m}{n}ln 2)}\] \[ln~p= -\frac{m}{n}(ln 2)^2\] \[m=-\frac{n ln~p}{(ln2)^2}\]
\[\frac{m}{n}=-\frac{n \log_2p}{ln2}\approx-1.44\log_2p\] Therefore, the optimal hash functions required \[h=-1.44\log_2p\]

\begin{figure}[ht]
\centering
\includegraphics[width=0.4\textwidth]{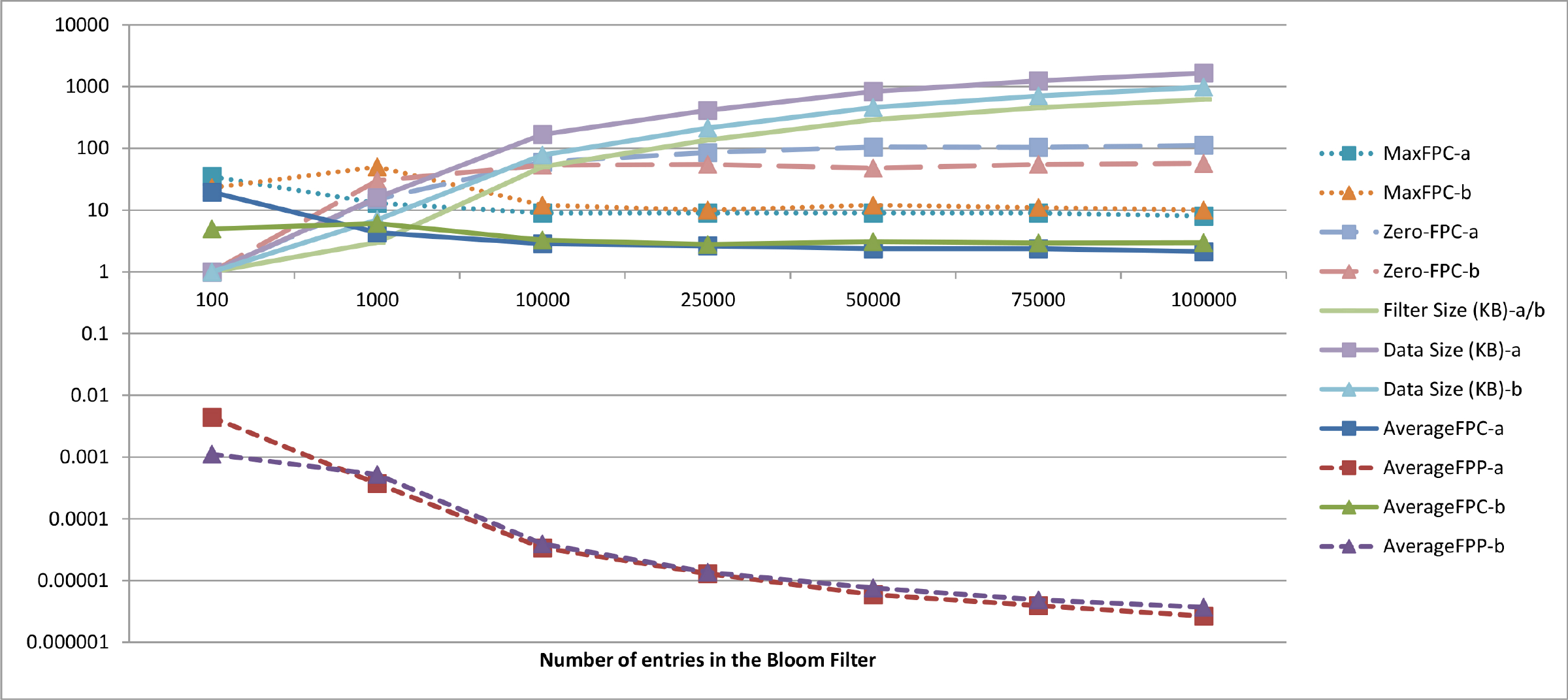}
\caption{Statistics on various data size during 1000 round queries.}
\label{s1}
\end{figure}

\begin{figure}[ht]
\centering
\includegraphics[width=0.4\textwidth]{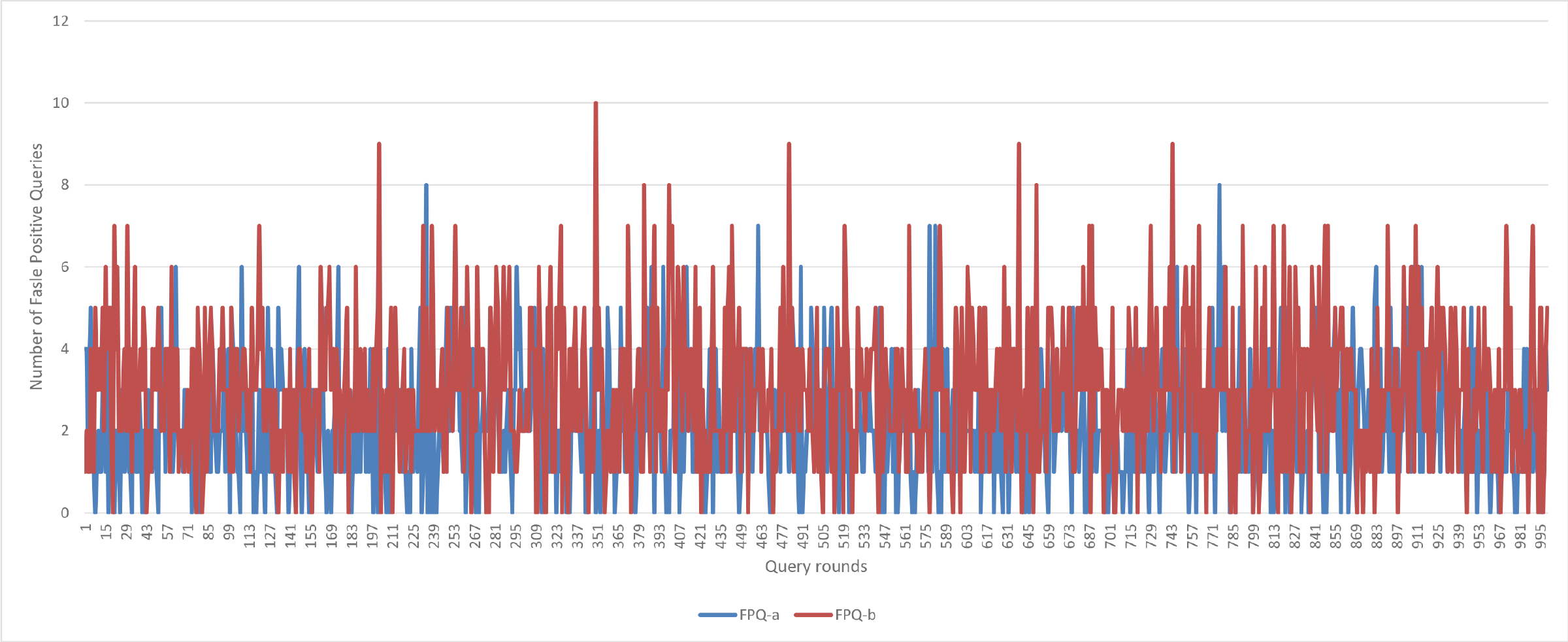}
\caption{False positive queries found on input size on 100000 elements during 1000 round queries. X-axis represent the number of query round and Y-axis represent the number of false positive queries.}
\label{s2}
\end{figure}

Figure~\ref{s1} depicts the experiments on various data size with 1000 round of queries \cite{Arash}. We have generated random string dataset of various size and combination of strings dataset of various size. The Table~\ref{tab1} describes the parameters of Figure~\ref{s1}. Figure~\ref{s1} represents the dynamic scaling of Filter Size according to data size. In Table~\ref{tab1}, the MaxFPC refers to a maximum number of false positive detected in 1000 round queries. The zero FPC refers to total number of zero false positive count in 1000 round queries. The AverageFPC and AverageFPP are the mean false positive count and the mean false positive probability in 1000 round queries respectively.

\begin{table}[ht]
\centering
\caption{Parameters description of Figure ~\ref{s1}}
\begin{tabular}{p{2cm}p{6cm}}
\hline
Name & Description \\ \hline
$a$ & Represents random strings dataset\\
$b$ & Combination of strings dataset\\
MaxFPC & Maximum number of false positive count in 1000 round queries \\ 
Zero FPC & Total number of no false positive count (Not found FP) in 1000 round queries \\ 
AverageFPC & $\frac{Total~FPC}{1000}$ \\
AverageFPP & $\frac{Total~FPP}{1000}$ \\
Filter Size & The Bloom Filter array size. \\
Data Size & Total number of input entries. \\ \hline

\end{tabular}
\label{tab1}
\end{table}

Figure~\ref{s2} depicts the snapshot by keeping the number of input to 100000 elements \cite{Arash}. The experiment is conducted by fixing the number of input elements in random string and combination of alphabets. Those strings are input to study the behavior of a number of false positive queries hit  in 1000 round queries. The dataset $a$ and dataset $b$ consist of random string and combination of the alphabets to form strings in different sizes. The input elements vary from 100 elements to 100000 elements to study the behavior of the false positive queries and the probability of false positive.  

\subsection{Discarding Compressed Bloom Filter} 
The Compressed Bloom Filter (ComBF) \cite{Mitz} reduces the extra space requirements, and maps an element into a single bit. However, there is an important tradeoff between performance and space complexity. Therefore, the ComBF does not exhibit a good performance.

\begin{figure}[ht]
\centering
\includegraphics[width=0.4\textwidth]{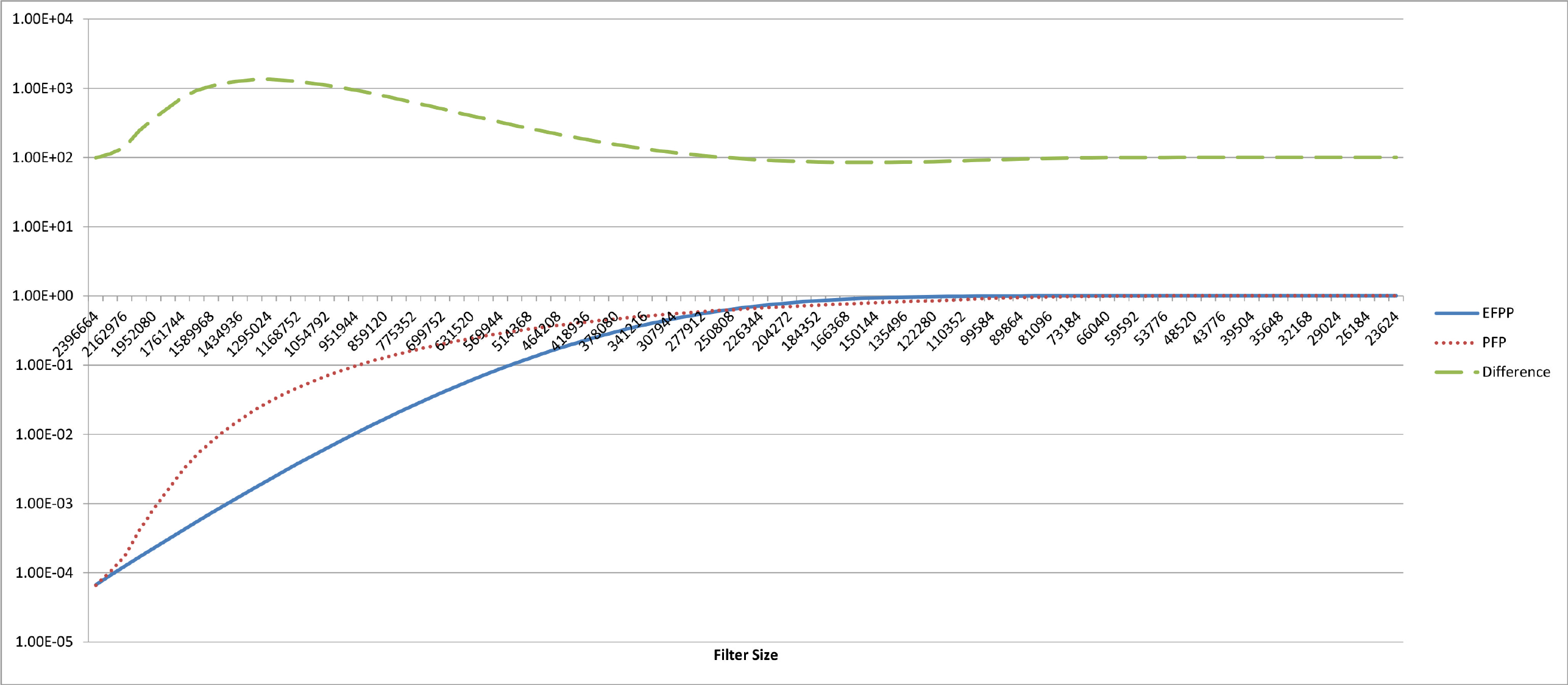}
\caption{Compression vs false positive on input of 100000 random string}
\label{comp}
\end{figure}

Figure ~\ref{comp} exposes the trade off between the compression and false positive. The high compression results in increment of false positive probability. In Figure ~\ref{comp}, the input strings are generated randomly and input to the compressed Bloom Filter. Hence, the result shows that high compression rate increases false positive which is not desirable for Bloom Filter. Figure ~\ref{comp} Effective False Positive Probability (EFPP), Probability of False Positive (PFP), and difference between both (Difference) \cite{Arash}. The difference is calculated as follows \[Differnce=100\times \frac{PFP}{EFPP}\]

\section{Variants of Bloom Filter}
\label{var}

\subsection{Scalable Bloom Filter} 
Scalable Bloom Filter (SBF) \cite{Almeida} is a Bloom filter having one or more Bloom Filters. In each Bloom filter, the array is partitioned into $k$ slices. Each hash function produces one slice. During insertion operation, for each element $k$ hash functions produces an index in their respective slice. So, each element is described using $k$ bits. When one Bloom filter is full, another Bloom filter is added. During query operation, all filters are searched for the presence of that element. The $k$ bit description of each element makes this filter more robust where no element is especially sensitive to false positives. In addition, this Bloom Filter have the advantage of scalability by adapting to set growth by adding a series of classic Bloom filters and making the error probability more tighter as per requirement.

\subsection{Adaptive Bloom Filter} 
Adaptive Bloom Filter (ABF) \cite{ABF} is a Bloom Filter based on the Partial-Address Bloom Filter \cite{peir}. ABF is used in tracking the far-misses. Far-misses are those misses that is hits, if the core is permitted to use more cache. Each set of each core, a Bloom Filter array (BFA) with $2^k$ bits is added. When a tag is removed from the cache, tag's $k$ least significant bit is used to index a bit of the BFA, which is 1. During Cache miss, using the $k$ least significant bit the BFA is looked for the requested tag. A far-miss is detected when the array bit becomes 1.

\subsection{Blocked Bloom filter} 
Blocked Bloom Filter \cite{Putze} is a cache-efficient Bloom Filter. It is implemented by fitting a sequence of $b$ standard Bloom Filter in each cache block/line. Usually for better performance, the bloom filters are made cache-line-aligned. When an element is added, the first hash function determines the Bloom filter block to use. The other hash functions are used for mapping of the element to $k$ array slots, but within this block. Thus, this lead to only one cache miss. This is further improved by taking a single hash function instead of $k$ hash functions. Hence, this single hash function determines the $k$ slots. In addition, this hash operation is implemented using fewer SIMD instructions. The main disadvantage in using one hash function is, two elements are mapped to same $k$ slots causes a collision. And this leads to increased false positive rate (FPR).

\subsection{Dynamic Bloom Filter} 
Dynamic Bloom Filter (DBF) \cite{DyBF} is an extension of Bloom Filter which changes dynamically with changing cardinality. A DBF consist of some CBF (Counting Bloom Filter), say $s$. Initially, $s$ is 1 and the status of CBF as active. A CBF is called active, when a new element is inserted or an element is deleted from it. During insertion operation, DBF first checks whether the active CBF is full. If it is full a new CBF is added and its' status is made active. If not, then new element is added to active CBF. During query operation, the response is given after searching all CBF. And, during deletion operation, first the CBF is found which contains the element. If a single CBF contains that element, then it is deleted. However, if multiple CBFs are there, then the deletion operation is ignored but deleted  response (i.e. the operation is completed) is delivered. Furthermore, if the sum of two CBF capacities is less than a single CBF then they are united. For that, addition of counter vectors is used. The time complexity of insertion is same i.e. $O(1)$, whereas query and deletion operation is $O(k\times s)$ where $k$ is the number of hash functions. 

\subsection{Deletable Bloom Filter} 
Deletable Bloom filter (DlBF) \cite{DBF} is a Bloom Filter that enables false-negative-free deletions. In this Bloom Filter, the region of deletable bits is encoded compactly and saved in the filter memory. DlBF, divide the Bloom Filter array into some regions. This region is marked as deletable or non-deletable using bitmap of size same as the number of regions. During insertion operation, when an element maps to an existing element slot, i.e. collision, then the corresponding region is marked as non-deletable i.e bitmap is assigned value 1. This information is used during deletion. The elements under deletable region are only allowed to be deleted. Insertion and query operations in DBF are same as the traditional Bloom Filter.

\subsection{Index-Split Bloom Filters} 
Index-split Bloom filter (ISBF) \cite{Split} helps in reducing memory requirements and off-chip memory accesses. It consist of many groups of on-chip parallel CBFs and a list of off-chip items. When a set of items is stored, the index of each item is divided into some $B$ groups. Each group contains $b$ bits, where ${B=\lceil log_2  n/b\rceil}$. So the items are split into $2^b$ subsets. Each subset is represented by a CBF. Thus, total $2^b$ CBFs per group are constructed in on-chip memory. During query operation, after matching the query element and the index of an item found by the $B$ group of on chip parallel CBFs, response is given. Also, for deletion operation, a lazy algorithm is followed. Because, deletion of an item requires adjustment of indexes of other off-chip items and reconstruction of all on-chip CBFs. Moreover, the average time complexity for off-chip memory accesses for insertion, query and, deletion is O(1).

\subsection{Quotient filter} 
Quotient Filter (QF) \cite{Bender} is a Bloom Filter where each element is represented by a multi-set $F$. The $F$ is an open hashtable with a total buckets of $m$=$2^q$, called quotienting\cite{knuth}. Besides, $F$ stores $p$-bit fingerprint for each element which is the hash value. In this technique, a fingerprint $f$ is partitioned into $r$ least significant bits, which stores the remainder. The $q$=$p$-$r$ is the most significant bits which stores the quotient. Both quotient and remainder is used for reconstruct of the full fingerprint. During insertion operation, $F$ stores the hash value. During query operation, $F$ is searched for the presence of the hash value of the element. And, during deletion operation, the hash value of that element is removed from $F$. QF has the advantage of dynamical resizing i.e. it expands and shrunk as elements are added or deleted. However, the QF insertion throughput deteriorates towards the maximum occupancy.

\subsection{NameFilter} 
Name Filter \cite{NameFilter} is a two-tier filter which helps in looking up names in Named Data Networking. The first tier determines the length of the name prefix and second tier makes use of the prefix determined in the previous stage to make a search in a group of Bloom Filters. In the first stage the name prefixes are mapped to Bloom Filter. Thereafter, the process of building up a Counting Bloom Filter is taken up. This filter is built for the concerned prefix set and then it is converted to take the form of a conventional Boolean Bloom Filter. As a final step, the second stage use the merged Bloom Filter. In the first stage, the calibration of the name prefixes to the Bloom Filter is done on the basis of their lengths. It maps the $k$ hash function into a single word. Hence, the Bloom Filter is called One Memory Access Bloom Filter as the query access time is $O(1)$ instead of $O(k)$. First, it acquires the hash output of the prefix using the DJB hash method. Then, the later hash value is calculated using the previous hash value. Thus, after $k-1$ loops, it obtains a single hash value and stores it in a word. This value is input for the calculation of the address in Bloom Filter, and the rest bits are calculated from one AND operation. So, when $k-1$ bits are 1s, then a graceful identification is declared. The aim of this stage is to find the longest prefix. In second stage, the prefixes are divided into groups based on their associated next-hop port(s). All groups are stored in the Bloom Filter. And, the desired port is found in this stage. In MBF, each slot stores a bit string with machine word-aligned. The $N$th bit stores the $N$th Bloom Filter's hash value and rest bits are padded with 0s. To obtain the forwarding port number, AND operation is done on $K$ bit strings with respect to $k$ hash functions. The location of 1 in the result gives the port number.

\subsection{Cuckoo Filter} 
A Cuckoo Bloom Filter \cite{Cuckoo} is based on Cuckoo hash table \cite{Cuckooh}. This Bloom Filter stores fingerprint instead of key-value pairs. Whereas, fingerprint means the hash value of the element. For insertion, index for two candidate buckets are calculated. One is the hash value of the element and another is the XOR operation between the hash value of the element and the hash value of the fingerprint of that element. This is called partial-key cuckoo hashing. This method reduces hash collision and improves the table utilization. After finding the indexes, the element is stored in any free bucket. otherwise cuckoo hash tables' kicking \cite{Cuckooh} of elements is done. For query operation, two candidate buckets are calculated as done in insertion operation, then if the element is present in any one of them true is returned otherwise false. For deletion operation, same procedure as lookup is followed, whereas instead of returning true or false, element is deleted. The advantage of the basic algorithms (i.e. insertion,deletion and lookup) is they are independent of hash table configuration (e.g. number of entries in each bucket). However, the disadvantage of using partial-key cuckoo hashing for storing fingerprints leads to slow increase in fingerprint size to increase in filter size. In addition, if the hash table is very large, but stores short fingerprints then hash collision increases. This lead to the chances of insertion failure and also reduces the table occupancy. 

\subsection{Multi-dimensional Bloom Filter} 
Crainiceanu et. al. proposed a Bloom Filter called Bloofi  \cite{Adina}. Bloofi is a Bloom Filter index. It is implemented like a tree. The Bloom Filter tree construction is done as follows. The leaves are Bloom Filters. And, the bitwise OR on the leaf Bloom Filters is done to obtain the parent nodes. This process continues till root is obtained. During lookup operation, the element is checked at root if it does not match then it returns false. Because if an element in leaf does not match then it will not match from the leaf to the root. Whereas, if the element matches, the query further moves to roots' children Bloom Filters till it reaches the leaf. During insertion of a new node, search for most similar node to the new node is done. As Bloofi want to keep similar nodes together. So, when found this new node is inserted as its sibling. If an overflow occurs, then the same procedure is followed as in a B+ tree. During deletion operation, the parent node deletes the pointer to the node. And, when underflow occurs, the same procedure is followed as in B+ tree.

\subsection{Sliding Bloom Filter} 
Sliding Bloom Filter \cite{Naor} is a Bloom Filter having a sliding window. It has parameters ($n$, $m$, $\varepsilon$). The sliding window remains over last $n$ elements and the value of the slots is 1. In other words, the window only shows the elements that are present. The $m$ numbers of elements that appear before the window elements does not have restrictions on the value. And $\varepsilon$ is the at most probable of slot being 1. This Bloom Filter is a dictionary based and use the Backyard Cuckoo hashing \cite{Backyard}. To this hashing a similar lazy deletion method is applied as used by Thorup \cite{thorup}. A parameter $c$ is used, which is the trade off between the accuracy of the index stored and the number of elements stored in the dictionary. After optimizing the parameter $c$ the Sliding Bloom filter shows good time and space complexity. The algorithm uses a hash function selected from the family of Universal hash functions. For each element in the dictionary $D$, stores its hash value and location where it previously appeared. The stream of data is divided into generations of size $n$/$c$ each, where $c$ is optimized later. Generation 1 is the first $n$/$c$ elements; generation 2 is next $n$/$c$ elements and so on. Current window contains last $n$ elements and at most $c$+1 generations. Two counters are used, one for generation number (say $g$) and another for the current element in the generation (say $i$). For every increment of $i$, $g$ get incremented to mod ($c$+1). For insertion, first obtain the $i$th hash value and checks whether it is present in $D$, if exists, the location of the element is updated with the current generation. Otherwise, it stores the hash value and generation number. Finally, update the two counters. If $g$ changes, then scan  $D$ and delete all elements with associated data equal to the new value of $g$.

\subsection{Bloom Filter Trie} 
Bloom Filter Trie (BFT) \cite{Trie} helps to store and compress a set of colored k-mers, and efficient traversal of the graph. It is an implementation of the colored de Bruijn graph (C-DBG). It is based on burst trie which stores k-mers along with the set of colors. Colors are bit array initialized with 0. A slot assigns the value 1 if that index k-mer has that color. Later, this set of color is compressed. BFT is defined as $t=(V_t,E_t)$ having the maximum height as $k$ where the k-mers is split into $k$ substrings. A BFT is a list of compressed containers. An uncompressed container of a vexter V is defined as $<s,color_{ps}>$ where $s$ is the suffix and $p$ is the prefix which represents the path from root to V. Tuples are ordered lexicographically based on their suffixes. BST support operations for storing, traversing, and searching of a pan-genome. And, it also helps in extracting relevant information of the contained genomes and subsets. The time complexity for insertion of a k-mer is $O(d+2^\lambda+2q)$ where $d$ is the worst lookup time, $\lambda$ is the number of bits to represent the prefix and $q$ is the maximum number of children. And, the time complexity of lookup operation is $O(2^\lambda+q)$.

\subsection{Autoscaling Bloom Filter} 
Autoscaling Bloom Filter \cite{Denis} is a generalization of CBF, which allows adjustment of its capacity based on probabilistic bounds on false positives and true positives. It is constructed by binarization of the CBF. The construction of Standard Bloom Filter is done by assigning all nonzero positions of the CBF as 1. And, given a CBF, the construction of ABF is done by assigning all the values which are less than or equal to the threshold value as 0.

\subsection{d-left Counting Bloom filter} 
d-Left CBF (dlCBF) \cite{d-left} is an improvement of the CBF. It uses the d-left hash table. This hash table consists of buckets, where each bucket has fixed number of cells. Each cell is of fixed size to hold a fingerprint and a counter. This arrangement makes the hash table appear as a big array. Each element has a fingerprint. And each fingerprint has two parts. The first part is a bucket index, which stores the element. Second part is the remainder part of the fingerprint. The range of bucket index is [$B$] and the remainder is [$R$]. So the hash function is $H$: $U$ $\rightarrow$ [$B$] $\times$ [$R$]. During element insertion, hash the element and store in appropriate remainders in the cell of each bucket. And increment the counter. And during deletion, decrement the counter. dlCBF solves the problems arise due to use of a single hash function. The hashing operation has two phases. In the first phase, apply a hash function, which gives the true fingerprint. And in the second phase, find the $d$ locations of the element using additional (pseudo)-random permutation. One small disadvantage in the obtained $d$ locations is, these are not independent, and uniform and as it is determined by the choices of the permutation.

\subsection{Ternary Bloom Filter} 
Ternary Bloom Filter (TBF) \cite{Lim} is another improvement of CBF. This Bloom filter introduces another parameter $v$ for each hash value, which can have possible values as {0, 1, $X$}. During insertion operation, if an element is mapped to a hash value for the first time, assign value 1 to $v$. If another element is mapped to the same hash value, then assign value $X$ to $v$. During lookup operation, if an element's every $v$ value for each hash value is $X$ then it is defined as indeterminable. Indeterminable means, the element cannot be identified as negative or positive. And, value 1 indicates, the element is present and value 0 indicates the element is absent. Similarly, in deletion operation, if an elements' every $v$ value for each hash value is $X$ then it is defined as undeletable. Undeletable means, the element cannot be deleted from TBF. And, if $v$ is value 1 it assigns value 0. TBF allocates the minimum number of bits to each cell which saves memory. In addition, it also gives much lower false positive rate compared to the CBF, when the same amount of memory used by both filters.

\subsection{Difference Bloom Filter} 
Difference Bloom Filter (DBF) \cite{Yang} is a probabilistic data structure based on Bloom Filter. It has multi-set membership query which is more accurate and has a faster response speed. It is based on two main design principles. First, to make the representation of the membership of elements exclusive by writing a different number of 0s and 1s in the same filter. Second, use of DRAM memory to increase the accuracy of the filter. DBF consist of a SRAM and a DRAM chaining hash table. The SRAM filter is an array of $m$ bits with $k$ independent hash functions. During the insertion function, elements in the set $i$ are mapped to $k$ bit of the filter. Arbitrarily $k-i+1$ bits are set to value 1 and other $i-1$ bits are set to value 0. This is called $<i,~k>$ constraint. If the new element gets conflicted with another element in the filter, DBF use dual-flip strategy to make this bit shared. Dual-flip is to change a series of mapping bits of the filters, so that the filter satisfy the $<i,~k>$ constraint. During lookup operation, if exactly $k-i+1$ bits are 1 then it returns true. During deletion operation, for each bit of the $k$ bits of an element, DBF decides whether to reset it or not with the help of DRAM table.  

\subsection{Self-adjustable Bloom Filter} 
TinySet \cite{TinySet} is a Bloom Filter that has more space efficiency compared to standard Bloom Filter. Its' structure is similar to the blocked Bloom filter. Whereas, each block is a chain based hash table \cite{rivest}. It uses a single hash function, $H$  $\rightarrow$ $B$ $\times$ $L$ $\times$ $R$, where $B$ is the block number, $L$ is the index of the chain within that block, and $R$ is the remainder (or fingerprint) that is stored in that block. 

All operations (insertion, deletion and lookup) initially follow three common steps. First, apply hash function to the element and obtain the $B$, $L$, and $R$ values. Second, use $B$ to access the specific block. Third, calculate the Logical Chain Offset (LCO) and Actual Chain Offset (ACO) values. During insertion operation, shift all fingerprints from offset to the end of the block to right. The first bits in a block contain a fixed size index ($I$). Unset $I$ means chain is empty. If the $I$ bit is unset, it is made to set and the new element is marked as the last of its chain. During deletion operation, if $I$ is unset, then the operation is terminated as it indicates the element is absent. Otherwise, shift all bits from the ACO value to end of the block to left by a single bit. If the deleted element is marked last then previous is made last or mark entire chain as empty. In lookup operation, if $I$ is unset similarly the operation is terminated. Otherwise, search the chain. TinySet is more flexible due to its ability to dynamically change its configuration as per the actual load. It accesses only a single memory word and partially support deletion of elements. However, delete operation gradually degrade its space efficiency over time.

\subsection{Multi-stage Bloom Filter} 
BloomFlow \cite{Craig} is a multi-stage Bloom Filter which is used for multicasting in Software-defined networking (SDN). It helps to achieve reductions in forwarding state while avoiding false positive packet delivery. The BloomFlow extends the OpenFlow \cite{openflow} Forward action with a new virtual port called BLOOM$\_$PORTS to implement Bloom filter forwarding. When a flow specifies an output action to BLOOM$\_$PORTS forwarding Element (FE) implements an algorithm. The algorithm first reads from the start of the IP option field, the Elias gamma encoded filter length $b$, and the number of hash function of $k$ fields. Then the algorithm treats the rest of the bits of the IP option field as a Bloom Filter. And, this Bloom Filter is copied to a temporary cache for further processing. The remainder of IP options fields and the IP payload are shifted left to remove first stage filter from the packet header. Then the algorithm  iterates through all interfaces and check for membership test for each interface's bloom identifier in the cached bloom filter. Bloom identifier is a unique, 16 bit integer identifier. The Bloom identifier is assigned by the network controller to every interface on the network that participates in multicast forwarding. If the membership test returns true, the packet is forwarded from the matched interface.

\subsection{Dynamic Reordering Bloom Filter} 
Dynamic Reordering Bloom Filter \cite{DChang} is another type of Bloom Filter that saves the searching cost of Bloom Filter. It dynamically reorders the searching sequence of multiple Bloom Filter using One Memory Access Bloom Filter (OMABF) and the order of checking is saved in Query Index (QI). This approach considers two factors. First, policy of changing the query priority of Bloom Filter. Second, reduction of overhead caused due to change in the order. This approach reduces the searching time of the query by sorting and saving the query data in Bloom Filter based on popularity. Sorting is done based on the query order, i.e popularity of data. So when the request comes from that data it quickly gives the response. And, when the popularity of a data becomes more, its query order is made a level higher in the Bloom Filter. However, this change of query order imposes overheads. To solve this, Query Index (QI) is used. QI saves the query priority of each block. When membership is checked Bloom Filter are checked according to the order saved in QI.

\section{Conclusion}
\label{con}
The Bloom Filter is the widely used data structure. The Bloom Filter also associates with a system to improve the performance dramatically. Moreover, it does not waste more spaces of main memory. The Bloom Filter provides a fast lookup system with a few KB of memory spaces. The Bloom Filter returns either 0 (False) or 1 (True). However, this Boolean value is classified into four categories, namely, TN, TP, FP, and FN. The TN and TP boost up the lookup performance of a system. On the contrary, the FP, and FN become an overhead to the system. Nevertheless, the FN is not common for all variants of Bloom Filter. The FP is the key barrier of Bloom Filter. Therefore, there are several kinds of Bloom Filters in the market. The key objective of the modern Bloom Filter is to reduce the probability of FP. In addition, the modern Bloom Filter also deals with high scalability, space efficiency, adaptability, and high accuracy. Besides, the Bloom Filter meets copious applications, and thus, extensive experiment has been done on Bloom Filter. The paper discusses a few selected applications to highlight the efficacy of the Bloom Filter. However, it is observed that the Bloom Filter is applied extensively in computer networking. Moreover, the efficiency, and accuracy of Bloom Filter depends on the probability of false positive. Therefore, reducing the false positive probability is a prominent challenge to achieve. Finally, the Bloom Filter will be able to reduce the false positive probability approximately to zero.

In this paper, we presented the theoretical and practical analysis of Bloom Filter. Moreover, there are abundant of Bloom Filter variants, those are discussed in this paper. Also, we have exposed the disadvantages of compressed bloom filter through a experiment. Moreover, the FP analysis is also shown through an experiment. 
\balance
\bibliographystyle{IEEEtran}
\bibliography{mybibfile}

\end{document}